\newcommand{\be}{\begin{equation}}\newcommand{\ee}{\end{equation}}
\newcommand{\bea}{\begin{eqnarray}}\newcommand{\eea}{\end{eqnarray}}
\newcommand{\nn}{\nonumber}\newcommand{\p}[1]{(\ref{#1})}
\newcommand{\lb}[1]{\label{#1}}
\newcommand\s{\scriptscriptstyle}

\newcommand\q{\quad}
\newcommand\qq{\quad\quad}
\renewcommand\={\ =\ }
\newcommand\Tr{\mbox{Tr}\,}

\newcommand\cA{{\cal A}}

\newcommand\cD{{\cal D}}
\newcommand\cC{{\cal C}}

\newcommand\cN{{\cal N}}

\newcommand\cW{{\cal W}}

\newcommand\olp{\overleftarrow{\partial}}
\newcommand\orp{\overrightarrow{\partial}}

\newcommand\olQ{\overleftarrow{Q}}
\newcommand\orQ{\overrightarrow{Q}}

\newcommand\tpa{\theta^{+\alpha}}

\newcommand\tma{\theta^{-\alpha}}

\newcommand\tka{\theta^{\alpha}_k}

\newcommand\tib{\theta^{\beta}_i}

\newcommand\btka{\bar\theta^{\da k}}
\newcommand\btkb{\bar\theta^{\db k}}

\newcommand\btpa{\bar{\theta}^{+\dot{\alpha}}}

\newcommand\tp{\theta^+}
\newcommand\tm{\theta^-}
\newcommand\btp{\bar\theta^+}

\def\a{\alpha}
\def\da{{\dot\alpha}}
\def\b{\beta}
\def\db{{\dot\beta}}
\def\g{\gamma}

\def\d{\delta}

\def\eps{\epsilon}

\def\ve{\varepsilon}

\def\bph{{\bar\phi}}
\def\vp{\varphi}

\def\j{\psi} \def\bj{{\bar\psi}}

 \def\th{\theta}

\def\si{\sigma}

\def\J{\Psi}
\def\bJ{\bar\Psi}
\def\L{\Lambda}

\def\pa{\partial}
\def\na{\nabla}

\newcommand\ada{{\alpha\dot{\alpha}}}

\newcommand\ab{{\alpha\beta}}
\newcommand\ba{{\beta\alpha}}

\newcommand\pada{\partial_{\alpha\dot{\alpha}}}

\newcommand\A{{\s A}}

\newcommand\C{{\s C}}
\newcommand\R{{\s R}}

\newcommand\sL{{\s L}}

\newcommand\W{{\s W}}
\newcommand\Z{{\s Z}}

\newcommand{\pp}{{\s ++}}
\newcommand{\m}{{\s --}}

\newcommand{\Dp}{D^{\pp}}

\newcommand{\dpp}{\partial^{\pp}}

\newcommand{\Vp}{V^\pp}
\newcommand{\Vm}{V^\m}

\newcommand{\Dpa}{D^+_\alpha}

\newcommand{\bDpa}{\bar{D}^+_{\dot{\alpha}}}

\def\sfrac#1#2{{\textstyle\frac{#1}{#2}}}
\def\ha{\frac12}
\def\sha{\sfrac12}

\def\e{\mbox{e}}

\def\diff{\mbox{d}}

\documentclass[12pt]{article}
\usepackage{amsmath,amssymb}

\begin{document}

\begin{titlepage}

\begin{center}
{\Large\bf
Non-Anticommutative Deformations of \\
\vspace{0.3cm}

N=(1,1) Supersymmetric Theories}
\vspace{1.5cm}

{\large\bf
E.A. Ivanov,   B.M. Zupnik}\\
\vspace{1cm}

 {\it Bogoliubov  Laboratory of Theoretical Physics, JINR,
141980 Dubna, Russia} \\
{\tt eivanov, zupnik@thsun1.jinr.ru}

\end{center}
\vspace{1cm}

\begin{abstract}
\noindent We discuss chirality-preserving nilpotent deformations 
of four-dimensional
$N{=}(1,1)$ Euclidean harmonic superspace and their implications
in $N{=}(1,1)$ supersymmetric gauge and hypermultiplet theories, 
basically following {\tt [hep-th/0308012]} and {\tt [hep-th/0405049]}. 
For the SO(4)$\times$SU(2) 
invariant deformation, we present non-anticommutative Euclidean analogs 
of the $N{=}2$ gauge multiplet and hypermultiplet off-shell actions. 
As a new result, we consider a specific non-anticommutative hypermultiplet 
model with $N{=}(1,0)$ supersymmetry. It involves 
free scalar fields and interacting right-handed spinor fields.

\end{abstract}
\vskip3cm
\begin{center}
{\it Submitted to Proceedings of the Seminar ``Classical and Quantum
Integrable Systems'' dedicated to the memory of M.V. Saveliev (Dubna,
Russia, January 26-29, 2004)}
\end{center}

\end{titlepage}

\setcounter{page}{1}
\section{Introduction}
In recent years, non-(anti)commutative deformations of supersymmetric
field theories received a great deal of attention.

The simplest type of non-commutativity affects the 
space-time  coordinates
\be
x^m\star x^n-x^n\star x^m=i\Theta^{mn} \lb{A}
\ee
where  $\Theta^{mn}$ is some constant tensor specifying
the deformation. Such non-commutative coordinates arise
in the field-theory limit of string theory in a constant
$B$-field background \cite{SW,DN}. For local fields $f(x)$ and $g(x)$,
this non-commutativity implies the use of the Moyal-Weyl star-product
which can be defined via the bi-differential operator $P$ (Poisson 
structure)
\be
f\star g=f e^Pg,\q P={i\over2}\Theta^{mn}\olp_m\orp_n\,.\lb{fg}
\ee

Moyal-Weyl type deformations of supersymmetric theories in superspace
are characterized by a generic Poisson bracket $APB$
where $A$ and $B$ are some superfields and the Poisson operator $P$
is in general some quadratic form in derivatives with respect to
both the even and odd superspace coordinates \cite{FL,KPT}.
Symmetry properties of the operator
$P$ determine unbroken symmetries of the deformed superfield theory:
these symmetries are those generators of which commute with $P$.
\footnote{In general, this criterion should be applied in a weak sense,
i.e. for the commutator sandwiched between the superfields
$A$ and $B$.}

The specific deformed superfield field theories studied so far correspond
to some particular degenerate choices of the generic superdifferential 
Poisson operator $P$. E.g., the authors of \cite{BS} considered the 
deformations of some theories in harmonic $N=2$ superspace 
\cite{GIK1,GIOS} corresponding to the standard pure bosonic Poisson 
structure \p{fg}.

Deformations of a different kind are the nilpotent or non-anticommutative
ones for which the operator $P$ is bilinear in the proper derivations 
with respect to Grassmann coordinates. As such one can choose  either 
generators of supersymmetry (Q-deformations), or spinor covariant 
derivatives (D-deformations). A surge of interest in superfield theories 
deformed in such a way was triggered by a recent paper~\cite{Se} where a 
minimal deformation of the Euclidean $N{=}(\ha,\ha)$ superspace
was considered. For the chiral $N{=}(\ha, \ha)$ coordinates
$(x^m_\sL,\theta^\alpha,\bar\theta^\da)$ the operator $P$ defining the
relevant star product is given by the simple bracket
\be
APB\=-\sha(-1)^{p(A)}C^\ab\pa_\a A\pa_\b B\,, \q P = 
-\sha C^\ab \olp_\a \orp_\b\,, \lb{P1}
\ee
with $C^\ab$ being some constant symmetric matrix,  
$\pa_\a=\partial/\partial \th^\a$, and $p(A)$ the $Z_2$-grading.
The operator $P$ defined by \p{P1} acts on the $\theta^\alpha$ 
coordinates only and retains the $N{=}(\ha,0)$ fraction of the original
$N{=}(\ha,\ha)$ supersymmetry. It is very important that the 
corresponding noncommutative product of superfields preserves the chiral
and antichiral representations of the $N=(\ha,\ha)$ supersymmetry.
Like the bosonic deformation \p{A}, \p{fg}, this purely fermionic
deformation also originates from string theory, as discussed in
\cite{Se} and \cite{BrSc}-\cite{vN}.

Deformations of the $N{=}2$ superfield theories along similar lines
were discussed in \cite{FLM}. In this contribution we shall focus on 
the harmonic-superspace formalism of the nilpotently deformed Euclidean 
$N{=}(1,1)$ theories, basically following Refs. \cite{ILZ,FS,FILSZ} 
(see also \cite{AIO,AI}).

The  Grassmann harmonic analyticity is the key notion of the off-shell 
superfield description of $N{=}2$ supersymmetric field theories
in four dimensions~\cite{GIK1,GIOS} where it plays the role analogous to
chirality in $N=1$ superfield theories. In particular, the analytic gauge
and hypermultiplet superfields  are the building-blocks of off-shell
interactions, and the harmonic analytic superspace formalism is 
indispensable for quantum supergraph calculations. By construction, the 
nilpotent Q-deformations (and some special D-deformations) of $N{=}(1,1)$ 
Euclidean superspace  preserve this harmonic G-analyticity \cite{ILZ,FS}. 
Yet, the chirality also plays the important role in $N{=}2$ and 
$N{=}(1,1)$ supersymmetric gauge theories, so the deformations which we 
shall consider preserve  as well both chiralities.

In Section 2 we review the nilpotent Q-deformations of the Euclidean chiral
$N{=}(1,1)$ superspace and analyze the role of the standard conjugation
or an alternative pseudoconjugation in Euclidean $N{=}(1,1)$ 
supersymmetric theories. The corresponding bi-differential operator~$P$  
preserves chirality and anti-chirality, and half of the original
$N{=}(1,1)$ supersymmetry ($N{=}(1,0)$ supersymmetry).
For special choices, however, $N{=}(1,\ha)$ supersymmetry or
the whole automorphism group SO(4)$\times\,$SU(2) can be retained.

Section 3 is devoted to the chirality-preserving SO(4)$\times\,$SU(2)
invariant deformation of the gauge $N{=}(1,1)$ theories in the harmonic
superspace. This singlet deformation breaks half of supersymmetries
and gives rise to some additional
interactions of the scalar field $\bph$ of the $N=(1,1)$ gauge
multiplet with the remaining components of the latter \cite{FILSZ}.
\footnote{The singlet Q-deformation of U(1) gauge theory was 
independently considered in \cite{AI}.} 

Non-anticommutative interactions of the Grassmann-analytic hypermultiplets
are considered in Section~4. Formally these interactions resemble those
considered in the bose-deformed harmonic superspace of~\cite{BS},
however, the component contents of these two theories
are entirely different. As a new explicit  example, we analyze
in some detail the simplest hypermultiplet self-interaction which
vanishes in the anticommutative-superspace limit. In the component
action of this model, the scalar fields do not interact with fermions, 
and only some specific fermionic self-interaction is present, 
with two derivatives on fermions. The solvable equation
for the right-handed fermions contains the nonlinear source constructed from
the left-handed ones which are free.

\setcounter{equation}0
\section{Deformations of N=(1,1) Euclidean  chiral \break superspace}
The Euclidean $N{=}(1,1)$ superspace has as its automorphisms
the Euclidean space spinor group $Spin(4)\sim$ SU(2)$_\sL\times$ SU(2)$_\R$
and the R-symmetry group SU(2)$\times\,$O(1,1) properly acting on
the coordinates $ x^m, \tka, \btka$.
We prefer to use the chiral coordinates $z_\sL\ \equiv\ ( x^m_\sL, \tka,
\btka )$ to parametrize this superspace.
These Euclidean coordinates $z_\sL$ are real with respect to the standard
conjugation \cite{Zu}
 \be
 \widetilde{\tka}\=\varepsilon^{kj}\varepsilon_\ab\theta_j^\beta\ ,\qq
\widetilde{\btka}\=-\varepsilon_{kj}\varepsilon_{\da\db}\bar\theta^{\db j}
\ ,\qq
\widetilde{AB}\=\widetilde{B}\widetilde{A}\ .
\lb{ELconj}
\ee
This conjugation squares to identity on any object, and with respect to it
the  $N{=}(1,1)$ superspace has the real dimension $(4|8)$.
 However, if we wish to treat the $N{=}(\ha,\ha)$ superspace as a real
subspace of the $N{=}(1,1)$ superspace (like $N=1$ supersubspace in
the standard Minkowski $N=2, 4D$ superspace), e.g.  in order to be able
to make reductions to the theories considered in \cite{Se}, we cannot
limit ourselves merely to this standard conjugation.
Indeed, the Euclidean $N{=}(\ha, \ha)$ superspace
{\it cannot} be real with respect to the complex conjugation: two
independent SU(2) spinor coordinates have the real dimension 8
which coincides with the Grassmann dimension of the whole $N{=}(1,1)$
superspace.

The alternative SU(2)-breaking pseudoconjugation in the same  Euclidean
$N{=}(1,1)$ superspace was considered in \cite{ILZ}:
\be
(\tka)^*\=\varepsilon_\ab\theta^\beta_k\ ,\qq
(\btka)^*\=\varepsilon_{\da\db}\bar\theta^{\db k}\ ,\qq
(x^m_\sL)^*\=x^m_\sL\ ,\qq
(AB)^*\=B^* A^*.
\lb{sr2conj}
\ee
The existence of this pseudoconjugation does not impose
any further restriction on the $N{=}(1,1)$ superspace
which has the same dimension $(4\vert 8)$
as with respect to the complex conjugation. Clearly, with respect to this
pseudoconjugation, $\theta^\alpha_1$ and $\bar\theta^{\dot\alpha 1}$
are `real', so they form an $N{=}(\ha, \ha)$ subspace of the `real'
dimension $(4\vert 4)$ in $N{=}(1,1)$ superspace (such subspaces
can be singled out in a few different ways). The standard conjugation 
\p{ELconj} and the
pseudoconjugation \p{sr2conj} act differently on the objects transforming by
non-trivial representations of the R-symmetry  SU(2).\footnote{Some 
ambiguities of generalized conjugations in Grassmann algebras 
($C$-antilinear maps with  squares equal to $\pm 1$) 
were discussed in \cite{Ma}.}
The map ${}^*$ squares to $-1$ on the Grassmann coordinates and the
associated spinor fields, and to $+1$ on any bosonic monomial or field.
On the singlets of SU(2), both maps act as the standard complex conjugation.
In particular, the invariant actions are real with respect to both
${}^*$ and ${}^\sim$, despite the fact that the component fields may
have different properties under these (pseudo)conjugations.

After this digression, let us come back to our main subject,
Q-deformations of $N{=}(1,1)$ theories.
In chiral coordinates, the simplest Poisson structure
operator is
\be
P = -\frac{1}{2}C^{\alpha\beta}_{ik}\olQ^i_\alpha\orQ^k_\beta
= -\frac{1}{2}C^{\alpha\beta}_{ik}\olp^i_\alpha\orp^k_\beta
\ee
and the Poisson bracket for two superfields $A$ and $B$ is defined as
\bea
&&A P B = -\sha(-1)^{p(A)}(\partial^k_\alpha A)C^\ab_{kj}(\partial^j_\beta B)
= -(-1)^{p(A)p(B)}B P A\ .
\lb{Poper}
\eea
Here, $C^\ab_{kj}=C^\ba_{jk}$ are some constants,
$\;p(A)$ is the $Z_2$-grading,
and the partial spinor derivatives act as
\be
\partial^k_\alpha\tib=\delta_i^k\delta^\beta_\alpha
\qq\textrm{and}\qq
\bar\partial_{\da i}\btkb=\delta^k_i\delta^\db_\da\ .
\ee
By definition, the bracket \p{Poper} preserves both chirality and
anti-chirality and does not touch SU$(2)_R$ acting on dotted indices.
Generically, it breaks half of the original $N{=}(1,1)$ supersymmetry
since the generators $\bar Q_{\dot\alpha k}$ do not commute with
the operator $P$. We demand $P$ to be real, i.e.~invariant under some antilinear map
in the algebra of superfields.
The two possible (pseudo)conjugations  lead to
different conditions on the constants $C^\ab_{kj}$.
The constant deformation matrix can be split into
two irreducible parts,
\be
C^\ab_{kj}\=C^{(\ab)}_{(kj)}+2\varepsilon^\ab\varepsilon_{kj}I\,,
\lb{const}
\ee
where $I$ is a real parameter.
The second, singlet part preserves the full
SO$(4)\times\,$SU(2) symmetry:
\be
P_s\ = -I\,\overleftarrow{Q}^k_\alpha \overrightarrow{Q}^\alpha_k\,, \quad
AP_s B\=-I(-1)^{p(A)}Q^k_\alpha A Q^\alpha_k B\,. \lb{Cinv}
\ee

Given the operator \p{Poper}, the Moyal product of two superfields reads
\bea
A \star B &=& A\,\e^P B \=
A\,B+ A\,P\,B + \sha A\,P^2 B + \sfrac16 A\,P^3 B + \sfrac{1}{24} A\,P^4 B
\lb{moyal}
\eea
where the identity $P^5=0$ was used. This star product preserves
both chirality and antichirality and breaks $N{=}(0,1)$ supersymmetry.
In the approach with the star product only free actions preserve all
supersymmetries while interactions get deformed and they are not
invariant under the $N{=}(0,1)$ supersymmetry transformations.

The (3,3) part $C^{(\alpha\beta)}_{(kl)}$ of the deformation matrix
breaks the R-symmetry SU(2), so we should choose one of the
alternative reality conditions to define the minimal form of the
matrix $C^{(\ab)}_{(kl)}$.
The  minimal representation of this (3,3) part has the following form:
\bea
&&C^{(\ab)}_{(12)} \=C^{(\ab)},\q C^{(\ab)}_{(11)}=C^{(\ab)}_{(22)}
=0\,,\nn\\
&&AP_\C B=-\frac12(-1)^{p(A)}C^{(\ab)}(Q^1_\alpha A Q^2_\beta B+Q^2_\alpha A
Q^1_\beta B)\,,
\eea
if we assume that $C^{(\alpha\beta)}_{(ik)}$ is real 
with respect to the $\widetilde{\;}$ conjugation, 
$\widetilde{C^{(\alpha\beta)}_{(ik)}}= C_{(\alpha\beta)}^{(ik)}\,$.

The choice of the ${}^*$ pseudoconjugation \p{sr2conj}
is compatible with the decomposition of $N{=}(1,1)$ into two
$N{=}(\ha,\ha)$ superalgebras.
Therefore, it allows one to choose a degenerate deformation
\be
P(Q^2)
\=-\sha C (\overleftarrow{Q}{}^2_1\,\overrightarrow{Q}{}^2_2+
\overleftarrow{Q}{}^2_2\,\overrightarrow{Q}{}^2_1)\,,\lb{1break}
\ee
which does not involve $Q^1_\alpha$ and contains the real parameter $C$.
In this case, only $\bar Q_{\da 2}$ are broken, but not the supercharges
$\bar Q_{\da 1}$. Hence, the deformation $P(Q^2)$
preserves the larger fraction $N{=}(1,\ha)$ of the original $N{=}(1,1)$
supersymmetry.

It is of course possible to consider more general deformations affecting
both the chiral and anti-chiral sectors. E.g. one can take the
anticommuting set of pseudoreal generators $Q^2_\alpha, \bar Q_{\da 1}$
and construct the real deformation operator $\hat{P}$ and the corresponding
bracket for even superfields $A$ and $B$ as
\be
A\hat{P}B=-C^\ab Q^2_\alpha AQ^2_\beta B-B^{\alpha\da}
(Q^2_\alpha A\bar Q_{\da 1}B+\bar Q_{\da 1}AQ^2_\alpha B)-\bar C^{\da\db}
\bar Q_{\da 1}A\bar Q_{\db 1}B.\lb{hatP}
\ee
It is evident that this deformation operator defines an associative
 star-product and it commutes with all spinor derivatives $D^k_\alpha,
 \bar D_{\da k}$, as well as with  4
generators of supersymmetry $Q^2_\alpha, \bar Q_{\da 1}$. Hence
it breaks half of supersymmetry and preserves both chiralities.

\setcounter{equation}0
\section{Chirality-preserving singlet deformations of \break
N=(1,1) harmonic superspace}

Harmonic superspace with noncommutative bosonic coordinates $x^m_\A$ has been
discussed in~\cite{BS}. This deformation yields nonlocal theories
but preserves the whole $N{=}2$ supersymmetry. The nilpotent D-deformations
of Euclidean $N{=}(1,1)$ superspace also preserving the full amount
of supersymmetry were considered in \cite{FLM}. Within the harmonic superspace
formalism, a special case of such deformations,
the singlet one
preserving the SO(4)$\times$SU(2) symmetry, one of two chiralities
and harmonic analyticity, was addressed in \cite{ILZ,FS}. In
particular, in \cite{FS} $N{=}(1,1)$ gauge theory
with such D-deformation was studied (see also a recent preprint \cite{Ke}).
Further in this contribution we shall not discuss
this type of nilpotent deformations. Instead, we shall concentrate
on the supersymmetry-breaking singlet nilpotent Q-deformation
associated with the operator $P_s$ \p{Cinv}. We shall essentially use
the Euclidean version of the harmonic superspace approach,
following refs. \cite{ILZ,FILSZ}.

The basic concepts of the harmonic superspace approach in its
Euclidean variant coincide, up to a few minor
distinctions, with those of the standard (Minkowski)
$N{=}2, D{=}4$ harmonic superspace as collected in the
book \cite{GIOS}. In both versions, the key ingredient is the SU(2)/U(1)
harmonics $u^\pm_i, \quad u^{+i}u^-_i = 1$, 
where SU(2) is the R-symmetry group. 
The chiral-analytic coordinates $Z_\C=(x^m_\sL,\theta^{\pm\alpha},
\bar\theta^{\pm\da}, u^\pm_i) $ in the  $N{=}(1,1)$ harmonic superspace
are related to the analytic coordinates via the shift of the bosonic
coordinate
\bea
&& x^m_\A =
x^m_\sL-2i(\sigma^m)_\ada\theta^{-\alpha}\btpa\,, \quad
\theta^{\pm \alpha} = \theta^{\alpha i}u^\pm_i\,,\;
\bar\theta^{\pm \dot\alpha} = \bar\theta^{\dot\alpha i}u^\pm_i\,.
\lb{Ccoor}
\eea
The (pseudo)conjugations  \p{ELconj} and \p{sr2conj}
can be extended to the harmonics and the coordinates of the harmonic
superspace \cite{ILZ}. These two (pseudo)conjugations act identically
on invariants and harmonic superfields,
e.g.~$(A^kB_k)^*=\widetilde{(A^kB_k)}$ or
$(q^+)^*=\widetilde{q^+}$, but they differ when acting
on harmonics or R-spinor component fields, e.g.~$(A_k)^*\neq\widetilde{A_k}$.
An important invariant pseudoreal subspace
is the analytic Euclidean harmonic superspace,
parametrized by the coordinates
\be
(x_\A^m~,~\theta^{+\alpha}~,~\bar\theta^{+\da},u^\pm_k ) \ \equiv\
(\zeta, u)\,. \lb{ChirAn}
\ee
The supersymmetry-preserving spinor and harmonic derivatives
in different coordinate bases are defined in \cite{ILZ,FS,FILSZ}.
A Grassmann-analytic ($G$-analytic) superfield $\Phi=\Phi(\zeta,u)$
is defined by the constraints
\be
\Dpa \Phi(\zeta, \theta^-, \bar\theta^-, u) \=
\bDpa \Phi(\zeta, \theta^-, \bar\theta^-, u) \= 0.
\ee
It is important that the chirality-preserving operator $P$
\p{Poper}  also preserves  Grassmann analyticity:
\be
 [P,(\Dpa, \bDpa)] \=   0 \ .
\ee

In what follows it will be convenient to deal with harmonic projections of
the $N{=}(1,1)$ supersymmetry generators
\be
Q^k_\alpha=u^{+k}Q^-_\alpha-u^{-k}Q^+_\alpha,\q
\bar Q_{\da k}=u^+_k\bar Q^- -u^-_k\bar Q^+.
\ee
For instance, in the chiral-analytic coordinates we have
\be
Q^+_\alpha=\partial_{-\alpha},\q Q^-_\alpha=-\partial_{+\alpha}
\ee
where $\partial_{\pm\alpha}=\partial/\partial\th^{\pm\alpha}$.
In these coordinates, different terms in the  product
\p{moyal} with the singlet Q-deformation operator $P_s$
are explicitly expressed as
\bea
&&AP_sB=I(-1)^{p(A)}\,(\partial_{-\a} A \partial^\a_+B
+\partial^\a_+ A \partial_{-\a} B)\,,\nn\\
&&{1\over2}AP^2_sB=-{I^2\over4}(\pa_+)^2A(\pa_-)^2B-{I^2\over4}(\pa_-)^2A
(\pa_+)^2B
+I^2\pa_{+\b} \pa^\a_-A\pa_{+\a} \pa^\b_-B\,,\nn\\
&&{1\over6}AP^3_sB={I^3\over4}(-1)^{p(A)}\pa^\a_-(\pa_+)^2A \pa_{+\a}
(\pa_-)^2B
+{I^3\over4}(-1)^{p(A)}\pa_{+\a}(\pa_-)^2A \pa^\a_-(\pa_+)^2B\,,\nn\\
&&{1\over24}AP^4_sB={I^4\over16}(\pa_+)^2(\pa_-)^2A (\pa_-)^2
(\pa_+)^2 B\,. \lb{12}
\eea
Note that the last two terms vanish for the analytic superfields.

Now we turn to some details of the deformed
$N{=}(1,1)$ gauge theory in harmonic superspace. It largely
mimics the harmonic superspace formulation of non-abelian
$N=2$ gauge theory in 4D Minkowski space \cite{GIOS}.

The basic superfield of the $N=(1,1)$ gauge theory is the analytic
anti-Hermitian potential $\Vp$ with the values in the algebra
of the gauge group which we choose to be U($n$). The gauge transformation
of the U($n$) gauge potential $\Vp$ reads
\be
\delta_\L\Vp\=\Dp \L+[ \Vp,\L]_\star
\ee
where $\L$ is an anti-Hermitian analytic gauge parameter and $\Dp$,
in the chiral-analytic basis, is
\be
D^{++} = \partial^{++} + \theta^{+\alpha}\partial_{-\alpha}
+\bar\theta^{+\dot\alpha}\partial_{-\dot\alpha}\,, \quad
\partial^{++} = u^{+i}\frac{\partial}{\partial u^{-i}}\,.
\ee
In the Wess-Zumino ($WZ$) gauge  we shall use the expansion of the potential in $\btpa$
\bea
&&\Vp_{\W\Z} =\bph^{++}+\btp_\da V^{+\da}+(\btp)^2V\,,\nn\\
&&\bph^{++}(x_\A,\tp,u)= (\tp)^2\bar\phi,\q V^{+\da}(x_\A,\tp,u)
=2\tpa A_\a^\da+4(\tp)^2\bar\Psi^{-\da}\,,\nn
\\
&&V(x_\A,\tp,u)= \phi  +4\tpa \Psi^-_\alpha\ +3 (\tp)^2\cD^{--}
\lb{WZ}
\eea
where $\J^-_\a=u^-_k\J^k_\a,
\bJ^-_\da=u^-_k\bJ^k_\da, \cD^{--}=u^-_ku^-_l\cD^{kl}$ and
all component fields are functions of~$x^m_\A$.

For what follows it will be convenient to rewrite the expression
for the $WZ$-potential in the chiral-analytic basis,
using the relation \p{ChirAn}
\bea
&&\Vp_{\W\Z}(Z_\C,u)=v^{++}(z_\C,u)+\btp_\da v^{+\da}(z_\C,u)
+(\btp)^2v(z_\C,u)
\lb{WZchir}
\eea
where the chiral superfunctions depend on the coordinates
$x_\sL^m,\tpa, \tma $ and $u^\pm_i$ only
\bea
&&v^{++}(z_\C,u)=(\tp)^2\bph(x_\sL)\,,\nn\\
&& v^{+\da}(z_\C,u)=V^{+\da}(x_\sL,\tp,u)
-2i\tma\pa_\a^\da\bph^{++}(x_\sL,\tp,u)\nn\\
&&=-2\tp_\a A^\ada+
4(\tp)^2 u^-_k \bar\Psi^{\da k}+2i\tm_\a(\tp)^2\pa^\ada\bph\,, \nn\\
&&v(z_\C,u)=V(x_\sL,\tp,u)+i\tma\pada V^{+\da}(x_\sL,\tp,u)-(\tm)^2\Box
\bph^{++}(x_\sL,\tp,u)\nn\\
&&=\phi+4\tpa \Psi^-_\a +3 (\tp)^2\cD^{--}
-2i(\tp\tm)\pa_m A_m+\tp\si_{mn}\tm F_{mn}\nn\\
&&
+4i\tma(\tp)^2\pada\bar\Psi^{-\da}-(\tm)^2(\tp)^2\Box\bph\,.\lb{Vpchir}
\eea
Here all component fields (after separating the harmonic dependence)
are functions of $x_\sL^m$.

Now we specialize to the simplest case of the U(1) gauge group.
The  corresponding $P_s$-deformed gauge and $N{=}(1,0)$ supersymmetry
transformations of the component fields can be readily found 
\cite{FILSZ}. They are given, respectively, by
\bea
&&\delta_a\phi=-8IA_m\pa_m a~,\q\delta_a\bph=0\,,
\q\delta_aA_m=(1+4I\bph)\pa_m a\,,
\nn\\
 &&\delta_a\Psi^k_\a=-4I\bar\Psi^{\da k}\pada a\,,
\q \delta_a\bar\Psi^k_\da=0\,,\q
\delta_a \cD^{kl}=0\,\lb{defgauge}
\eea
and
\bea
&&\delta_\epsilon\phi=2\epsilon^{\a k}\Psi_{\a k}\,,
\q\delta_\epsilon\bph=0\,,\q
\delta_\epsilon A_m=\epsilon^{\a k}(\sigma_m)_\ada\bar\Psi_k^\da\,,\nn\\
&&\delta_\epsilon\Psi^k_\a=
-\epsilon_{\a l}\cD^{kl}+{1\over2}(1+4I\bph)(\si_{mn}\eps^k)_\a
F_{mn}-4iI\epsilon^k_\a A_m\pa_m\bph\,,\nn\\
&&\delta_\epsilon\bar\Psi^k_\da=-i\epsilon^{\a k}(1+4I\bph)\pada\bph\,,\nn\\
&&\delta_\epsilon \cD^{kl}=
i\pa_m[(\epsilon^k\si_m\bar\Psi^l+\epsilon^l\si_m\bar\Psi^k)
(1+4I\bph)]\lb{defsusy}
\eea
where $F_{mn}=\pa_mA_n-\pa_nA_m\,$.

The nonpolynomial superfield action of the Q-deformed gauge theory has been
given in \cite{ILZ} as an integral over the full superspace in the chiral
coordinates, by analogy with the undeformed $N{=}2$ superfield action
\cite{Z1}. It was shown in \cite{FILSZ} that the $P_s$-deformed U(1)
gauge action can be conveniently rewritten as the integral over
the chiral superspace
\be
S^{(I)}={1\over 4} \int d^4x_\sL d^4\theta  {\cal A}^2
\lb{153}
\ee
where $\cA(x_\sL,\tp,\tm,u)$ is the deformed  chiral
superfield strength. The latter appears as the lowest component
in the $\btpa$ expansion of the covariantly chiral
superfield strength $\cW$:
\be
\cW\equiv -\sfrac14(\bar D^+)^2\Vm=\cA+\btp_\da\tau^{-\da}
+(\btp)^2\tau^{-2} \lb{ExpW}
\ee
and the action \p{153} can be rewritten as
\be
S^{(I)}={1\over 4} \int d^4x_\sL d^4\theta  {\cal W}^2\,.
\lb{1533}
\ee
It can be shown that the remaining two components in \p{ExpW} do not
contribute to \p{1533}.

The composite harmonic connection $\Vm$ is connected with the
basic potential $\Vp$ via the deformed
harmonic zero curvature equation \cite{ILZ}
\be
D^{++}V^{--}-D^{--}V^{++}+[V^{++},V^{--}]_\star=0 \lb{zerocurv}
\ee
where, in the chiral-analytic basis,
$$
D^{--} = \partial^{--} + \theta^{-\alpha}\partial_{+\alpha}
+\bar\theta^{-\dot\alpha}\partial_{+\dot\alpha}\,, \quad
\partial^{--} = u^{-i}\frac{\partial}{\partial u^{+i}}\,.
$$

As a consequence of \p{zerocurv}, the chiral superfield $\cA$ satisfies the
homogeneous harmonic equation
\be
[\pa^\pp+(1+4I\bph)\tpa\pa_{-\a}]\cA=0 \lb{homcA}
\ee
and some additional nonlinear inhomogeneous equation \cite{FILSZ}:
\bea
&&[\pa^{++}+ (1+4I\bar\phi)\tpa\pa_{-\a}]\vp^\m
+2(\cA-v)
 -I\left(\pa^\a_{-}v^+_\da\pa_{+\a}v^{-\da}
-\pa_+^\a v^+_\da\pa_{-\a}v^{-\da}\right)
\nn\\
&&+ \,\frac{I^3}{4}\,\pa^\a_-(\pa_+)^2v^+_\da\pa_{+\a}(\pa_-)^2v^{-\da}
=0\lb{inhom}
\eea
where $v^{-\da}$ and $\vp^\m$ are the proper chiral coefficients 
of the expansion of $\Vm$ in $\bar\theta^{\pm \dot\alpha}\,$.
They can be calculated in terms of the component fields.

The undeformed chiral U(1) superfield strength has the following
component field content
\bea
&&W_0 (x_\sL, \theta^+, \theta^-,u)=\varphi+2\theta^+\psi^- -2\th^-\psi^+
+(\tp)^2 d^{--}\nn\\
&&-2(\tp\tm)d^{+-}+(\tm)^2d^{++}+
(\tm\sigma_{mn}\tp) f_{mn}\nn\\
&&+2i[(\tm)^2\th^+\sigma_m \pa_m\bar\psi^+ +(\tp)^2\th^-\si_m\pa_m\bar\psi^{-}]
-(\tp)^2(\tm)^2\Box\bph\, \lb{undef}
\eea
where $f_{mn}=\pa_ma_n-³a_ma_n,\q\psi^\pm_\a = \psi^i_\a(x_\sL) u^\pm_i,\q
 d^{+-}=u^+_ku^-_ld^{kl}(x_\sL)\,$,
etc. This superfield obeys the free harmonic equation
$
D^{++}W_0 = 0
$
and transforms under $N=(1,0)$ supersymmetry  as
\be
\delta_\epsilon W_0 = (\epsilon^{-\alpha}\pa_{-\a}
+\epsilon^{+\a}\pa_{+\a})W_0\,. \lb{standrule}
\ee

It is rather straightforward to show that $\cA$ can be constructed as
a nonlinear transformation of the  undeformed U(1) superfield strength
$W_0$
\bea
&&{\cal A}(x_\sL, \theta^+, \theta^-,u) = (1 +4I\bph)^2 W_0 (x_\sL, \theta^+,
(1+4I\bph)^{-1}\theta^-,u)\,.\lb{Rel}
\eea
The nonlinear relations between the undeformed and deformed U(1)
component fields following from \p{Rel} are
\bea
&&\vp=(1+4I\bph)^{-2}[\phi+4I(1+4I\bph)^{-1}(A^2_m+4I^2(\pa_m\bph)^2)]\,,
\nn\\
&&a_m=(1+4I\bph)^{-1}A_m,\q \bar\psi^k_\da=(1+4I\bph)^{-1}\bJ^k_\da\,,\nn\\
&&\psi^k_\a=(1+4I\bph)^{-2}[\Psi^k_\a+4I(1+4I\bph)^{-1}
A_\ada\bJ^{\da k}]\,,\nn\\
&&d^{kl}=(1+4I\bph)^{-2}[\cD^{kl}+8I\bJ_\da^k\bJ^{\da l}]\,.\lb{U1swrel}
\eea

The  $N{=}(1,0)$ supersymmetry transformation of the deformed chiral
superfield is given by
\bea
&&\delta_\epsilon \cA=[(1+4I\bph)\epsilon^{-\a}\pa_{-\a} +\epsilon^{+\a}
\pa_{+\a}]\cA\,.
\eea

The deformed U(1) gauge superfield action can be expressed in terms of
the abelian undeformed objects up to a total spinor derivative in the
integrand
\be
S^{(I)} = {1\over 4}\int d^4x_\sL d^4 \theta \, {\cal A}^2=
{1\over 4}\int d^4x_\sL d^4 \theta \,(1+4I\bph)^2W^2_0\,.\lb{U1Act}
\ee
Using the  redefinitions of the deformed fields \p{U1swrel},
one can obtain the component Lagrangian of the deformed U(1) gauge
theory as $L^{(I)}=(1+4I\bph)^2L_0$ where $L_0$ is the free undeformed
Lagrangian
\bea
&&L_0=-{1\over2}\vp\Box\bph+{1\over4}(f_{mn}^2+
{1\over2}\varepsilon_{mnrs}f_{mn}f_{rs})
-i\psi^\a_k\pada\bar\psi^{\da k}+{1\over4}(d^{kl})^2\,.
\eea
It is obvious that the scalar, fermionic and auxiliary terms in the
action can be given the form of the free
kinetic terms by properly rescaling the fields $\vp, \psi^k_\a$
and $d^{kl}$. However, the nonlinear interaction of the fields
$\bph$ and $f_{mn}\,$,
\be
{1\over4}(1+4I\bph)^2(f_{mn}^2+{1\over2}\varepsilon_{mnrs}f_{mn}f_{rs})\,,
\ee
cannot be removed by any field redefinition. 

Now let us shortly discuss how the above generalizes to the nonabelian
U($n$) case ($n \geq 2$). We use the WZ-gauge for the U($n$) potential \p{WZ},
and the corresponding deformed component gauge transformations are
\bea
&& \d_a\bph=-i[a,\bph],\q \d_r\bJ^k_\da=-i[a,\bJ^k_\da],\q\d_r {\cal D}^{kl}=
-i[a,{\cal D}^{kl}]\,,\nn\\
&&\d_a A_m=\pa_m a+i[A_m,a] +2I\{\bph,\pa_m a\}\,,\nn\\
&&\d_a\phi=-i[a,\phi] -4I\{A_m,\pa_m a\}-4iI^2[\Box a,\bph]\,,\nn\\
&&\d_a\J^k_\a=-i[a,\J^k_\a]-2I(\si_m)_{\a\dot\a}\{\bJ^{\dot\a k},\pa_m a\}.
 \lb{Ungauge}
\eea

The $P_s$-deformed U($n$) chiral gauge superfield $\cA$ satisfies
the following equation:
\bea
&&\Dp\cA+I\tpa\{\bph,\pa_{-\a}\cA\}+(\tp)^2[\bph,\cA]
+I^2[\bph,(\pa_-)^2\cA]=0\lb{harmcA}
\eea
where $\bph$ is the Hermitian matrix scalar field. It is convenient
to define the following
matrix operator:
\be
L=1+2I\{\bph,~~\}\,,\lb{Loper}
\ee
then the first two terms in eq.\p{harmcA} can be rewritten
as $(\dpp+L\tpa\pa_{-\a})\cA\,$.
The undeformed harmonic chiral U($n$) superfield $A$
has the following component expansion
\bea
A &=& \varphi+2\th^+ \psi^- -2\th^- \psi^+ +(\tp)^2 d^{--}
+(\tp\tm)([\vp,\bph]-2d^{+-})+(\tm)^2d^{++} \nn\\
&& +\, (\tp\si_{mn}\tm) f_{mn}+2(\tm)^2\th^+\left(i\xi^+ -[\bph,\psi^+]
\right)+2i(\tp)^2\th^-\xi^- \nn\\
&& -\,(\tp)^2(\tm)^2\left(p+[\bph,d^{+-}]\right)\lb{Unundef}
\eea
where all the component fields are $n\times n$ matrices and
the following short-hand notation is used:
\bea
&&\nabla_m=\pa_m
+i[a_m,~~~]\,,\q f_{mn}=\pa_ma_n-\pa_na_m+i[a_m,a_n]\,,\nn\\
&&\xi^k_\a=(\si_m)_\ada\na_m\bj^{\da k},\q p=\nabla_m^2\bph+
\{\bar\psi^{\da k},\bar\psi_{\da k}\}
+\frac12[\bph,[\bph,\vp]]\,.\lb{Acomp}
\eea

The deformed chiral U($n$) superfield can be written as a sum of two
$N{=}(1,0)$ covariant objects
\bea
&&\cA(x_\sL,\tp,\tm,u)=[L^2+L(1-L)(\tm \pa_-)-{1\over4}(1-L)^2(\tm)^2
(\pa_-)^2] A(x_\sL,\tp,\tm,u)
\nn\\
&&-4I^2\hat{A}(x_\sL,\tp,u)
 \lb{Unoper}
\eea
where $A$ is the undeformed U($n$) superfield \p{Unundef}, and 
  the  $\bph$-dependent matrix  operator  $L$ \p{Loper} 
  commutes with $\th^{\pm\a}$ and $\pa_{-\a}$ and acts on all matrix
quantities standing to the right. The second part
$\hat{A}$ is a traceless  chiral-analytic $N{=}(1,0)$ superfield
\bea
&&\hat{A}(x_\sL,\tp,u)=\hat{p}-[\bph,d^{+-}]
+2\tpa (i[\bph,\xi^-_\a]-[\bph,[\bph,\j^-_\a]])\nn\\
&&+(\tp)^2[\bph,[\bph,d^{--}]]\,,\q \hat{p}=p-\sfrac{1}{n}\Tr p\,.
\lb{hatA}
\eea
Both parts of $\cA$ are thus expressed in terms of the undeformed field
components of the superfield $A$ \p{Unundef}.

The $N{=}(1,0)$ supersymmetry transformation of $\cA$ has the following form:
\bea
&&\d_\eps\cA=2(\epsilon^-\theta^+)[\bph,\cA]+L\eps^{-\a}\pa_{-\a}\cA
+\eps^{+\a}\pa_{+\a}\cA\,.
\eea

It is worth noting that the undeformed anti-self-duality  equation
in the $N{=}(1,1)$ supersymmetric U($n$) gauge theory  \cite{DeO,Zu2}
can be written in the pure chiral superfield form as
\be
A=0\,,
\ee
which, as follows from \p{Unundef}, amounts to the following set of
matrix component equations
\bea
&&f_{mn}(\si_{mn})^\b_\a=0\,,\q \vp=\j^k_\a= d^{kl}=0\,,\nn\\
&&(\si_m)_\ada(\pa_m\bj^{\da k}+i[a_m,\bj^{\da k}])=0\,,\q 
(\nabla_m)^2\bph+\{\bj^{\da k}\,,
\bj_{\da k}\}=0\,.
\eea
These anti-self-dual U($n$) solutions preserve only the 
$N{=}(1,0)$ supersymmetry, so it is
natural that the same undeformed solutions survive in 
the $I$-deformed U($n$)
gauge theory
\be
A=0~\Leftrightarrow~\cA=0\,.
\ee

The $I$-deformed U($n$) gauge theory component action can be directly
obtained from the superfield chiral action
\be
{\cal S}_n=\frac14\int d^4x_\sL d^4\theta\,\Tr\cA^2=
\int d^4x_\sL d^4\theta\,\Tr\{\frac14 (LA)^2-2I^2\hat{A}A\}\,,
\ee
using relations \p{Unundef} and \p{hatA}. In the limit
$I\rightarrow 0$ the first term yields the action of the undeformed
U($n$) gauge theory. The non-standard second term contains higher derivative
terms, in particular  $I^2(\Box\bph)^2$, which can hopefully be removed
by a redefinition of the scalar field $\vp$ (so far we have checked
this only for the bilinear free part of the total action).

\setcounter{equation}0
\section{Interactions of hypermultiplets in deformed \break
harmonic superspace}

The free $q^+$  hypermultiplet actions of ordinary harmonic theory
\cite{GIOS} are not deformed in the non-anticommutative superspace:
\be
S_0(q^+)\= \frac12\int\!\diff u\,\diff\zeta^{-4}\ q^{+}_a\star \Dp q^{+a}
=\frac12\int\!\diff u\,\diff\zeta^{-4} q^{+}_a \Dp q^{+a}\,.
 \lb{freeAc}
\ee
Here $\diff\zeta^{-4}=\diff^4x_\A (D^-)^4$ and the additional 
`Pauli-G\"ursey' SU(2)$_P$ indices
$a, b=1, 2$ were introduced: $q^{+a}=\ve^{ab}q^+_b=(\tilde{q}^+, q^+)$ .
Let us consider the $\btpa$-expansion of the superfield doublet $q^{+a}$
in the analytic basis
\bea
q^{+a} &=& c^{+a}+\btp_\da \kappa^{\da a}+(\btp)^2b^{-a},\hspace{2cm} \nn\\
\Dp q^{+a} &=& \pa^{++}c^{+a}+\btp_\da(\pa^{++} \kappa^{\da a}+2i\tp_\a\pa^\ada
c^{+a}) \nn \\
&& +\, (\btp)^2(\pa^{++}b^{-a}+i\tpa\pada\kappa^{\da a}) \lb{qchir}
\eea
where
\bea
&&c^{+a}=f^a+\tpa \rho^a_\a+(\tp)^2g^a,\q
\kappa^{\da a}=\chi^{a\da}+\tpa r^{\da a}_\a+(\tp)^2\bar\Sigma^{a\da},\nn\\
&&b^{-a}=h^a+\tpa\Sigma^a_\a+(\tp)^2X^a \lb{COmp}
\eea
and, for brevity, the U(1) charges of the component fields 
$f^a,g^a,h^a,\ldots $ are suppressed. The component fields are 
functions of $x^m_\A$ and harmonics.
The chiral representation of the free action (i.e., with the integration over
$\bar\theta^{+\dot\alpha}$ manifestly performed) reads
\bea
S_0(q^{+}) &=& -\int du\, d^4x_\A\, d^2\tp [\sfrac12 b^{-a}\pa^{++}c^+_a
+\sfrac12c^{+a}\pa^{++}b^-_a +
\sfrac14\kappa^{\da a}\pa^{++}\kappa_{\da a}\nn\\
&&+\,\sfrac{i}{2}\tpa (c^{+a}
\pada\kappa^\da_a-\kappa^\da_a\pada c^{+a})]\,. \lb{freechir}
\lb{qchfree}
\eea

The non-anticommutativity shows up in the hypermultiplet self-interactions. If
we prefer to work in the manifestly SU(2)$_P$ covariant formalism, it is
convenient to define two independent combinations:
\be
\{q^{+a},q^{+b}\}_\star,\q \left[q^{+a},q^{+b}\right]_\star=
2q^{+a}P_sq^{+b}=\ve^{ab}\cC^{++}\,.
\lb{Self}
\ee
The square of the first superfield contracted with some 
SU(2)$_P$-breaking constant
parameter $C_{(ab)}$ gives a non-anticommutative generalization 
of the self-interaction $[q^{+a}q^{+b}C_{(ab)}]^2$ which yields the familiar 
Taub-NUT hyper-K\"ahler metric on the bosonic target space \cite{GIOS}.
Leaving this generalization for the future study, we shall consider 
a simpler example of the deformed self-interaction constructed out of 
the second combination in \p{Self} and vanishing in the anticommutative 
limit $I\rightarrow 0$
\be
S_\nu(q^+)\= -\frac{\nu}{4}\int\!\diff u\,\diff\zeta^{-4}\,
\cC^{++}\star\cC^{++}=
-\frac{\nu}{4}\int\!\diff u\,\diff\zeta^{-4}\,\cC^{++}\cC^{++}\lb{S4nc}
\ee
where $\nu$ is a coupling constant and the overall sign 
was chosen for further convenience. Note that this superfield interaction is
nilpotent, $(\cC^{++})^2\sim (\btp)^2$, and preserves both SU(2)$_P$ 
and the R-symmetry SU(2) which acts on harmonics.

One can easily calculate the chiral components of the composite superfields
\bea
\cC^{++} &=& q^+_aP_sq^{+a}=-4iI\btpa\pada q^+_a\pa^\a_+q^{+a}=
-4iI\btpa\pada c^+_a\pa^\a_+c^{+a}\nn\\
&&+\,2iI(\btp)^2(\pada\kappa^\da_a
\pa^\a_+c^{+a}-\pada c^+_a\pa^\a_+\kappa^{\da a})\,,\nn\\
(\cC^{++})^2 &=& -8I^2(\btp)^2B^+_\da B^{+\da},\q B^{+\da}(c^+)
=\pa^\ada c^+_a\pa_{+\a}
c^{+a}\,.\lb{calQ}
\eea
The deformed interaction \p{S4nc} contains  superfields $c^{+a}$
only
\bea
S_\nu^s(q^{+})&=& 2\nu I^2\int du\, d^4x_\A\, d^2\tp B^+_\da B^{+\da} \nn \\
&=& -2\nu I^2\int du\, d^4x_\A\, d^2\tp
\pada c^+_a\pa^\a_+c^{+a}\pa^{\b\da} c^+_b\pa_{+\b}c^{+b}\,.
\lb{chiract}
\eea

The total superfield action $S_0 (q^+)+ S_\nu (q^+)$ yields the hypermultiplet
equation of motion
\bea
\Dp q^{+a}= \nu q^{+a}P_s(q^+_bP_sq^{+b})
\equiv J^{(+3)a}(q^+) \lb{EqMot}
\eea
where $J^{(+3)a}(q^+)$ is the nonlinear nilpotent source. After performing
the $\theta$-integration, the total action contains an infinite number 
of auxiliary fields coming from the harmonic expansions of the components 
in \p{COmp}. These auxiliary fields can be
eliminated using the appropriate non-dynamical equations collected in the
$\theta^+, \bar\theta^+$ expansion of \p{EqMot}.

The non-dynamical equations of motion for $ c^{+a}$ and 
$\kappa^{\da a}$ have the following form:
\bea
\pa^{++}c^{+a}=0\,, \quad 
\pa^{++}\kappa_\da^a-2i\tpa\pada c^{+a}=0\,. \lb{Eqs12}
\eea
In components, the solution to these equations is given by
\bea
c^{+a}=u^+_kf^{ak}(x)+\tpa\rho_\a^a(x)\,,\quad
\kappa^a_\da=\chi^a_\da(x)+2iu^-_k\tpa\pada f^{ak}(x)\,.\lb{lineq}
\eea

The last equation, for the chiral component $b^{-a}\,$, 
also follows from eq. \p{EqMot}
\bea
\pa^{++}b^{-a}+i\tpa\pada\kappa^{\da a}
&=& -4\nu I^2[\pada c^{+a}\pa^{\b\da}\pa^\a_+c^{+b}\pa_{+\b}c^+_b-
\pa_mc^{+a}\pa_m
c^+_b(\pa_+)^2c^{+b}\nn\\
&&+\pa^\a_+c^{+a}\Box c^+_b\pa_{+\a}c^{+b}+\pa^\a_+c^{+a}
\pa^{\b\da}c^+_b\pada\pa_{+\b}c^{+b}] \lb{bequ}
\eea
and is solved by
\bea
b^{-a}=-4\nu I^2u^-_k[\pada f^{ak}(\pa^{\b\da}\rho^{\a b})\rho_{\b b}
+\rho^{\a a}\pa^{\b\da}f^k_b\pada\rho^b_\b+\rho^{\a a}\rho^b_\a\Box f^k_b
]
\eea
(eq.\p{bequ} involves also the set of dynamical equations for 
the physical fields $f^{ak}(x), \rho^a_\alpha(x)$ and 
$\chi^a_{\dot\alpha}(x)$; these equations can be re-derived from 
the on-shell action written in terms of the physical fields). 
Actually, $b^{-a}$ does not contribute to the total physical 
on-shell action  $S_0 + S_\nu $: the only place where it 
appears is the first two terms 
in \p{freechir}, and these terms vanish after employing first of eqs.\p{Eqs12} 
and integrating by parts with respect to $\partial^{++}$.  

Eliminating the auxiliary component fields from the action 
$S_0+S_\nu$ by the substitution \p{lineq},
one obtains the physical action of this model. It contains the standard
free kinetic terms for the physical bosonic and fermionic fields, as well as 
some fermionic self-interaction with two derivatives:
\bea
&&S=\int d^4x[\sha \pa_m f^{ak}\pa_mf_{ak}+ \sfrac{i}{2}\rho^{\a a}\pa_\ada
\chi^\da_a- \nu I^2(\pa_\ada\rho_{\g a})\rho^{\a a}(\pa^{\b\da}\rho^{\g b})
\rho_{\b b}]\,.
\eea

The scalar field $f^{ak}$ and the left-handed spinor field $\rho^a_\a$ satisfy
the free massless equations in this model
\bea
&&\Box f^{ak}=0,\q \pada\rho^{\a a}=0\,.
\eea
At the same time, the  equation for the right-handed spinor field
$\chi^a_\da$ contains the nonlinear spinor source depending on
the  left-handed spinor field
\bea
&&i\pada\chi^\da_a=
-4\nu I^2[\rho_{\b b}(\pa_\ada\rho_{\g a})(\pa^{\b\da}\rho^{\g b})
+\rho_{\b a}(\pa_{\g\da}\rho_{\a b})(\pa^{\b\da}\rho^{\g b})\nn\\
&&
+\rho_{\b b}(\pa_{\g\da}\rho^\g_a)(\pa^{\b\da}\rho^b_\a)+\rho^{\b b}\rho_{\b a}
\Box\rho_{\a b}]=-\nu I^2J_{\a a}[\rho(x)]\,.
\eea
Note that the last two terms in $J_{\a a}$ are vanishing 
on the mass-shell of the free fields $\rho^a_\a$.
The exact classical solution for $\chi^{\da a}$ is a sum of the free
right-handed fermion $\chi^{\da a}_0$ and the inhomogeneous solution with
the above nilpotent spinor source:
\bea
&&\chi^{\da a}=\chi_0^{\da a}+
i\nu I^2\int d^4y\pa^\ada_xD^0(x-y)J_\a^a[\rho(y)]\,,\\
&&\pada\chi_0^{\da a}=0,\q\Box_x D^0(x-y)=\delta^4(x-y).\nn
\eea
Thus the considered model is exactly solvable at the classical level.

The component form of some other nilpotently deformed 
$q^+$ self-interactions and the deformed hypermultiplet interactions 
with the analytic gauge superfield~ $\Vp$ will be studied elsewhere.

\section{Conclusions}
In this contribution, basically following refs. \cite{ILZ,FS,FILSZ}, 
we briefly reviewed recent results on the nilpotent non-anticommutative 
deformations of Euclidean $N{=}(1,1)$ superspace, with the main 
emphasis on the structure of the singlet Q-deformation of $N=(1,1)$ 
gauge theories. This deformation breaks half of $N=(1,1)$ supersymmetry, 
but preserves O(4) and SU(2) automorphism symmetries, as well as both 
chiralities and harmonic Grassmann analyticity. We also considered 
a simple new example of the Q-deformed hypermultiplet action, with the 
self-interaction vanishing in the anticommutative limit. This model 
is exactly solvable at the classical level.   

\section*{Acknowledgements}
Most of the results presented in the review part of this contribution 
were obtained in collaboration with S. Ferrara, O. Lechtenfeld and 
E. Sokatchev. We are grateful to them. This work was partially supported by 
the INTAS grant No 00-00254, grant DFG No 436 RUS 113/669-2, RFBR 
grant No 03-02-17440,NATO grant PST.GLG.980302 and a grant of the 
Heisenberg-Landau program.

\end{document}